\documentclass[twocolumn,         % Format : preprint, twocolumn
               showpacs,            % Pacs : showpacs, noshowpacs
               preprintnumbers,     % Preprint: preprintnumbers,
               aps,                 % Society: ...
               prd,          	    % Journal Style : pra, prb, prc, prd, pre,
               a4paper,             % Size : a4paper, ...
               superscriptaddress,      % Affiliation (Title) : groupedaddress,
               nofootinbib,         % Footnote: footinbib, nofootinbib
               tightenlines,        % Remove additional spaces in a line
               floats,floatfix
               ]{revtex4-1}              
\usepackage{graphicx}  % needed for figures
\usepackage{dcolumn}   % needed for some tables
\usepackage{bm}   
\usepackage{amsmath,amssymb}
\usepackage{soul}
\usepackage[caption=false]{subfig}
\usepackage{caption}
\usepackage{float}
\usepackage{makecell}
\usepackage[thinlines]{easytable}
\usepackage{array,booktabs}
\usepackage{lipsum} 
\usepackage[colorlinks=true,linkcolor=blue,citecolor=blue]{hyperref}

\usepackage{array,makecell}
\newcolumntype{P}[1]{>{\centering\arraybackslash}p{#1}}
\usepackage[total={7.3in, 10.0 in}]{geometry}
\usepackage[colorinlistoftodos]{todonotes} 
\begin{document} 

\title{Tracker behaviour of quintom dark energy and the Hubble tension}
\author{Nandan Roy}
\email{nandan.roy@mahidol.ac.th (Corresponding Author)} 
\affiliation{Centre for Theoretical Physics \& Natural Philosophy, Mahidol University, Nakhonsawan Campus, Phayuha Khiri, Nakhonsawan 60130, Thailand}

\author{L. Arturo Ure\~{n}a-L\'{o}pez} 
 \email{lurena@ugto.mx}
\affiliation{%
Departamento de F\'isica, DCI, Campus Le\'on, Universidad de
Guanajuato, 37150, Le\'on, Guanajuato, M\'exico}

\date{\today}

\begin{abstract}
We study the dynamics of the quintom dark energy model using state-of-the-art cosmological observations. The set of equations has been converted into an autonomous system using suitable transformations of the variables. We have discussed the fixed points of the model and the general phase-space behavior, in particular, in finding the existence of the tracker solutions for this model. The observations suggest that at late times the phantom field should dominate the dark energy sector with an approximately $15\%$ share to the quintessence counterpart, and with both fields tracking the background at early times. A Bayesian model comparison with $Lambda$CDM has also been done by computing the Bayes factor and a positive preference has been obtained for the quintom model. Although not fully resolved, the Hubble tension can be reduced to $2.6 \sigma$ when compared with the value of $H_0$ reported in~\cite{Riess:2021jrx} and to $1.6 \sigma$ when compared with that of~\cite{Scolnic:2023mrv}.
\end{abstract}

\maketitle
%\listoftodos

\section{Introduction}
Although the cosmological constant is the simplest and most consistent explanation for the accelerated expansion of the universe~\cite{Riess:1998cb, Perlmutter:1998np, Aghanim:2018eyx}, its constant nature gives rise to some famous problems in cosmology, such as the fine-tuning problem and the cosmological constant problem~\cite{Weinberg:1988cp, Peebles:2002gy, Padmanabhan:2002ji}. Furthermore, as cosmology enters the era of ultra-precision measurement, the constraint on some key cosmological parameters puts the cosmological constant in more challenges. Currently, the most significant tension in cosmology is the Hubble tension. There is a discrepancy between the measurements of the Hubble parameter from early universe data~\cite{Aghanim:2018eyx, Alam:2016hwk, Joudaki:2019pmv} and late-time data~\cite{Riess:2016jrr, Riess:2019cxk, Wong:2019kwg, Freedman:2019jwv} at the level of $5 \sigma$. For a more in-depth discussion of this topic, see~\cite{Sch_neberg_2022, Verde:2019ivm}. One solution to the problems associated with the cosmological constant is considering dynamical dark energy models that evolve over time~\cite{amendola2010dark, Bamba:2012cp, 2006IJMPD..15.1753C, 2006AIPC..861..179P}. The equation of state (EOS) for dark energy is fixed at $-1$ for the cosmological constant, whereas for dynamical dark energy models it varies over time. This has also been supported by the fact that the current estimated value of EoS of dark energy is $w_{DE} \leq -1$~\cite{Aghanim:2018eyx}.

Quite a large number of dynamical dark energy models have been proposed as alternatives to the cosmological constant. Scalar fields with different types of kinetic terms and potentials are often considered candidates of the dynamical dark energy~\cite{2006IJMPD..15.1753C} as, for example, quintessence, phantom, $K$-essence, etc.  Moreover, a wide variety of other models have been suggested that can reduce or alleviate Hubble tension, such as the early dark energy model~\cite{Karwal:2016vyq}, the running vacuum model~\cite{SolaPeracaula:2023swx,Rezaei:2022bkb,Rezaei:2019xwo,Rezaei:2021qwd}, phantom-crossing models~\cite{DiValentino:2020naf,Roy:2022fif}, etc.

For quintessence scalar field models in which a canonical scalar field is minimally coupled to gravity, the allowed value of the EOS of dark energy is $-1 \leq w_{DE} \leq 0$~\cite{Ratra:1987rm, Tsujikawa:2013fta}. A value of EOS $w_{DE} < -1$ is not allowed for quintessence models. The consideration of a phantom scalar field or $K$-essence models can allow us to achieve an EOS that is less than $-1$~\cite{Nojiri:2005sx,Caldwell:2003vq, Caldwell:1999ew,roy2018dynamical} but crossing the critical limit $w_{DE} = -1$ cannot also be achieved in these models. Studies have suggested that $w_{DE}$ crosses over at $-1$~\cite{gerardi2019reconstruction, johri2004phantom, perkovic2019transient,Roy:2022fif} which means that at present $w_{DE} \leq -1$ but in the past $w_{DE} > -1 $.  

In order to achieve a phantom barrier crossing, one possible solution is to incorporate both quintessence and phantom scalar fields as components of the universe. These phenomenological models are referred to as quintom models~\cite{guo2005cosmological, cai2010quintom, zhao2006quintom, setare2008coupled, zhao2005perturbations, setare2009quintom}. These models have been successful in replicating the phantom divider crossing behavior of $w_{DE}$~\cite{zhao2006quintom,panpanich2019resolving}, suggesting that quintom models may be able to address the Hubble tension. However, a comprehensive comparison with recent cosmological observations is necessary to confirm this. 

In this work, we have considered a general quintom model without any specific form of the potential. Instead, we have considered the general parameterization of the potentials proposed in~\cite{roy2018new} and later used in~\cite{roy2019arbitrariness,LinaresCedeno:2021aqk,Urena-Lopez:2020npg,Roy:2023uhc} to study quintessence and phantom models. We have constructed an autonomous system considering the appropriate transformation of variables. As shown in our previous work for the quintessence field~\cite{Urena-Lopez:2020npg} we have considered the polar transformation and for the phantom field, we have considered the hyperbolic transformation~\cite{LinaresCedeno:2021aqk}. Similarly to the previous work, we can identify the existence of the tracker solutions for the quintom field with four different scenarios. 

\textit{Tracker solutions} are important in alleviating scalar-field dark energy models from coincidence and fine-tuning problems~\cite{Ratra:1987rm,Steinhardt:1999nw}. These solutions have a remarkable property where the energy density of the scalar field follows the background energy density, behaves like an attractor solution for a wide range of initial conditions. We have used recent cosmological observations to constrain the cosmological parameters. To compare the performance of our model with the $\Lambda$CDM model, we have used the concept of Bayesian model selection by calculating the Bayes factor. Additionally, we have provided an update on the current status of the Hubble tension in this model.
% The dynamical systems analysis of the quintom models has already been performed from different perspectives\cite{setare2009quintom,mishra2018dynamical, cai2010quintom, marciu2019dynamical}. Here, we have performed the dynamical system analysis using the general parameterization of the potential mentioned above, which includes a large class of popular potentials. We have found the tracker conditions for the quintom model. Three tracking scenarios have been found when both the quintessence field and the phantom field track simultaneously, and one of the fields track the background. A detailed cosmological data analysis of the model has also been performed together with a Bayesian model comparison with $\Lambda$CDM. We have also reported the current status of the Hubble tension in our model.
 
The paper is structured as follows. Section~\ref{sec:math-back} deals with the mathematical setup of the quintom model with general parameterization of the potentials. Section~\ref{sec:critical} describes the critical points and the behavior of the tracker solutions and the estimation of the initial condition. The constraint on the cosmological parameters from the current observations is discussed in Section~\ref{sec:constraints}. We summarize our findings in Section~\ref{sec:conclusions}.

\section{Mathematical background \label{sec:math-back}}

A quintom dark energy model consists of two minimally coupled scalar fields: one scalar field is the canonical scalar field ($\phi$), named quintessence, and another is the phantom scalar field ($\sigma$). The action of such a scenario is given by
\begin{eqnarray}
S = \int d^{4}x \sqrt{-g} \bigg[\frac{1}{2\kappa^2} R +\cal{L}_\text{DE}
+ \cal{L}_\text{M}\bigg]~, \label{actionquint}
\end{eqnarray}
 with the DE Lagrangian $\cal{L}_\text{DE}$ given explicitly by
 \begin{equation}
\begin{split}
{\cal{L}_\text{DE}} = - \frac{1}{2} g^{\mu\nu}\partial_{\mu}\phi\partial_{\nu}\phi-V_{\phi}(\phi) \\ 
+ \frac{1}{2} g^{\mu\nu}\partial_{\mu}\sigma\partial_{\nu}\sigma-V_{\sigma}(\sigma) \, .
\end{split}
\end{equation}
Here, $\kappa^2\equiv 8\pi G$ is the gravitational coupling and $V_\phi(\phi)$, $V_\sigma(\sigma)$ are the respective potentials for the canonical and phantom field, respectively. The term $\cal{L}_\text{M}$ corresponds to the standard matter content of the universe. We consider all components of the universe to be connected by a barotropic equation of state (EoS) $\rho_i=w_i p_i$. For example, $w_i = 1/3$ for radiation and $w_i=0$ for matter.

\subsection{Quintom equations of motion}
Once we consider the universe to be flat, homogeneous, and isotropic, the Friedmann equations are written as follows,
\begin{subequations}
\label{eq:FR}
\begin{eqnarray}
H^{2}&=&\frac{\kappa^{2}}{3}\Big( \sum_i \rho_i+\rho_{\phi}+\rho_{\sigma}\Big) \, , \label{FR1} \\
\dot{H}&=&-\frac{\kappa^2}{2}\Big[ \sum_i (\rho_{i}+p_i)+\rho_{\phi}+p_{\phi}+\rho_{\sigma}+p_{\sigma}\Big] \, , \label{FR2}
\end{eqnarray}
\end{subequations}
where $H=\dot{a}/a$ is the Hubble parameter and $a$ is the scale factor of the universe. The respective Klein-Gordon equations of the two constituents of the quintom model are:
\begin{subequations}
\label{eq:motion}
\begin{eqnarray}
\ddot{\phi} + 3H\dot{\phi} + \frac{\partial V_\phi(\phi)}{\partial\phi} &=&0 \, , \label{eq:pddot} \\
\ddot{\sigma} + 3H\dot{\sigma} - \frac{\partial V_\sigma(\sigma)}{\partial\sigma} &=& 0 \, . \label{eq:sddot}
\end{eqnarray}
\end{subequations}

In the above expressions, $p_\phi$ and $\rho_{\phi}$ are, respectively, the pressure and density of the canonical scalar field, and similarly $p_\sigma$ and $\rho_\sigma$ are the pressure and density of the phantom field. Their mathematical expressions are given by:
\begin{subequations}
\begin{eqnarray}
 \rho_{\phi} = \frac{1}{2}\dot{\phi}^{2} + V_\phi(\phi) \, , \quad p_{\phi} = \frac{1}{2}\dot{\phi}^{2} - V_\phi(\phi) \, , \label{rhophi} \\
 \rho_{\sigma} = -\frac{1}{2}\dot{\sigma}^{2} + V_\sigma(\sigma) \, , \quad p_{\sigma} = - \frac{1}{2}\dot{\sigma}^{2} - V_\sigma(\sigma) \, . \label{rhosigma} 
\end{eqnarray}
\end{subequations}

Since in a quintom model, the DE sector is the combination of the canonical and phantom fields and there is no interaction considered, its total density and pressure can be written in the following way;
\begin{subequations}
\begin{equation}
\rho_{DE}\equiv\rho_\phi+\rho_\sigma \, , \quad p_{DE} \equiv p_\phi+p_\sigma \, , \label{eq:rhoDE}
\end{equation}
while its total equation of state is given by
\begin{equation}
w_{DE} \equiv\frac{p_{DE}}{\rho_{DE}} = \frac{p_\phi+p_\sigma}{\rho_\phi+\rho_\sigma} \, . \label{eq:wDE}
\end{equation}
\end{subequations}

Also, using the conservation equations, the total density can be written as
$\rho_{tot}\equiv \sum_i \rho_i+\rho_\phi+\rho_\sigma$, obtaining:
\begin{equation}
 \dot{\rho}_{tot}+3 H(1+w_{tot})\rho_{tot}=0 \, , \label{rhot}
\end{equation}
with the corresponding total EoS,
\begin{equation}\label{eq:wtot}
 w_{tot}=\frac{p_\phi+p_\sigma+\sum_i p_i}{\rho_\phi+\rho_\sigma+\sum_i \rho_i}=w_\phi\Omega_\phi+w_\sigma\Omega_\sigma+\sum_i w_i \Omega_i \, .
\end{equation}
The individual EoS parameters are defined as
\begin{eqnarray}
 w_\phi=\frac{p_\phi}{\rho_\phi} \, , \quad w_\sigma=\frac{p_\sigma}{\rho_\sigma} \, , \quad  w_i=\frac{p_i}{\rho_i} \, , \label{eq:EoSs}
\end{eqnarray}
together with the corresponding density parameters,
\begin{eqnarray}
 \Omega_\phi\equiv\frac{\rho_\phi}{\rho_{tot}}~,~~\Omega_\sigma\equiv\frac{\rho_\sigma}{\rho_{tot}}~,~~\Omega_i\equiv\frac{\rho_i}{\rho_{tot}}~.
\end{eqnarray}
Notice that the total DE density parameter is given by the combination $\Omega_\phi+\Omega_\sigma\equiv\Omega_{DE}$, whereas the combination of all density parameters results in the Friedmann constraint from Eq.~\eqref{FR1}: $\Omega_\phi+\Omega_\sigma+\Omega_m + \Omega_r=1$.

\subsection{Autonomous system of equations}
To write the system~\eqref{eq:motion} as a set of autonomous equations, we define the following set of dimensionless polar variables,\footnote{For references on these changes of variables of scalar fields see for instance~\cite{Copeland:1997et,roy2018dynamical,roy2018new,roy2019arbitrariness}.}
\begin{subequations}
  \label{var_phi}
  \begin{align}
    \frac{\kappa \dot{\phi}}{\sqrt{6} H} = \Omega_{\phi} ^{1/2} \sin(\theta_\phi /2) \, , \quad \frac{\kappa V_{\phi}^{1/2}}{\sqrt{3} H} = \Omega_{\phi}^{1/2} \cos(\theta_\phi /2) \, , \label{eq:phi.a} \\
    y_{1\phi} \equiv - 2\sqrt{2} \frac{\partial_{\phi} V_{\phi}^{1/2}}{H} \, , \quad y_{2\phi} \equiv - 4\sqrt{3} \frac{\partial^2_\phi V_{\phi}^{1/2}}{\kappa H} \, , \label{eq:phi.b}             
  \end{align}
\end{subequations} 
and hyperbolic ones,
\begin{subequations}
  \label{var_sig}
  \begin{align}
    \frac{\kappa \dot{\sigma}}{\sqrt{6} H} =& \Omega_{\sigma} ^{1/2} \sinh(\theta_\sigma /2) \, , \quad \frac{\kappa V_{\sigma}^{1/2}}{\sqrt{3} H} \, = \Omega_{\sigma}^{1/2} \cosh(\theta_\sigma /2) \, , \label{eq:si.a} \\
    y_{1\sigma} \equiv& - 2\sqrt{2} \frac{\partial_{\phi} V_{\sigma}^{1/2}}{H} \, , \quad y_{2\sigma} \equiv - 4\sqrt{3} \frac{\partial^2_\phi V_{\sigma}^{1/2}}{\kappa H} \, , \label{eq:si.b}             
  \end{align}
\end{subequations} 
where the variables with suffix $\phi$ represent the variables related to the canonical field and with $\sigma$ represent the phantom field. As a result, the field equations~\eqref{eq:motion} are rewritten as the following set of equations,
\begin{subequations}
\label{eq:pol}
  \begin{eqnarray}
    \theta_\phi^{\prime} &=& - 3 \sin \theta_\phi + y_{1\phi} \, , \label{eq:pol_a} \\
    y_{1 \phi}^{\prime} &=& \frac{3}{2} \gamma_{tot} y_{1\phi}
                     + \Omega_{\phi}^{1/2} \sin(\theta /2) y_{2 \phi}  \, , \label{eq:pol_b} \\ 
    \Omega_\phi ^{\prime} &=& 3 (\gamma_{tot} - \gamma_{\phi}) \Omega_\phi \, , \label{eq:pol_c} \\
    \theta_\sigma ^{\prime} &=& - 3 \sinh \theta_\sigma - y_{1 \sigma}  \, , \label{eq:pol_d} \\
    y_{1\sigma}^{\prime} &=& \frac{3}{2} \gamma_{tot} y_{1\sigma} + \Omega_{\sigma}^{1/2} \sinh(\theta_\sigma /2) y_{2 \sigma}  \, , \label{eq:pol_e} \\ 
    \Omega_\sigma ^{\prime} &=& 3 (\gamma_{tot} - \gamma_{\sigma}) \Omega_\sigma \, . \label{eq:pol_f}
  \end{eqnarray}
\end{subequations}

 From Eqs.~\eqref{eq:EoSs}, the barotropic EOS of the canonical scalar field can be written in terms of the polar variable as $\gamma_\phi = 1 +w_\phi = 1- \cos \theta_\phi $, and then $0 < \gamma_\phi < 2$, while that of the phantom field is $\gamma_\sigma = 1+ w_\sigma = 1 - \cosh \theta_\sigma$, from which $\gamma_\sigma < 0$. The barotropic EOS parameter of the total dark energy, see Eq.~\eqref{eq:wDE}, is given by
\begin{equation}
 \gamma_{DE} = \frac{\gamma_\phi \Omega_{\phi} + \gamma_\sigma \Omega_{\sigma}}{\Omega_\phi + \Omega_\sigma}  \, .\label{eq:QuintomEoS}
\end{equation}

%\subsection{General form of the quintom potentials}
One can see that the system of equations~\eqref{eq:pol} is not closed until one specifies the functional form of $y_{2 \phi}$ and $y_{2 \sigma}$. These two variables, by definition, carry information about quintessence and phantom potentials. In general, there are two different ways to consider the functional forms of the variables $y_2$. First, one can consider a particular form of potential and then calculate the corresponding $y_2$, and secondly, one assumes a particular form of $y_2$ in terms of variables $y$ and $y_1$, and integrate back to get the actual form of the potential.

In this work, we use the parameterization of $y_2$ proposed in~\cite{roy2018new}, which includes a large class of dark energy potentials (see Table~1 and~2 in~\cite{roy2018new}). Thus, we have for the quintom potential,
\begin{subequations}
\label{eq:GP}
\begin{eqnarray} 
y_{2 \phi} &=& \alpha_0 y_\phi + \alpha_1 y_{1\phi} + \alpha_2 y^2_{1 \phi}/y_\phi \, , \label{eq:GP1} \\
y_{2 \sigma} &=& \beta_0 y_\sigma + \beta_1 y_{1\sigma} + \beta_2 y^2_{1 \sigma}/y_\sigma \, , \label{eq:GP2}
\end{eqnarray}
\end{subequations}
where $\alpha$ and $\beta$ are just constant parameters.

As explained in~\cite{roy2018new}, the parameters $\alpha$ and $\beta$ are called active parameters of the potentials, as they are the only ones that appear explicitly in the equations of motion. There are other parameters in the potentials, but these are instead called passive parameters, and their values can be connected to the initial values of the dynamical variables $\theta_{\phi,\sigma}$, $y_{1\phi,\sigma}$ and $\Omega_{\phi,\sigma}$.

Though the above-mentioned parametrization covers a wide range of potentials, there are still some potentials which cannot be written in the above form. For example, $V(\phi) = \sum_{i}^{n} \Lambda_i \exp(\lambda_i \phi)$ \cite{copeland1999generalized, ivashchuk2003cosmological} and $V(\phi) = V_0 \tanh^2(\frac{\lambda_1 \phi}{m_p}) \cosh(\frac{\lambda_2 \phi}{m_p})$ \cite{Bag:2017vjp} are not included in the list. However, examining the parameterizations given in~\eqref{eq:GP} allows us to include a wide range of potentials in our system, which are frequently used in the relevant literature. 

\section{Critical points with tracking behavior \label{sec:critical}}
Here we discuss the critical points of the quintom model that are of physical interest, particularly with tracking behavior. This study will be useful for understanding the combined dynamics of the quintom field and our particular choice of initial conditions.

% \textbf{Luis: I reordered the subsections for a better presentation of the discussion on critical points. I also hid the figure on the phase space of the equations of state, as it is not very clear what our intention is exactly with the curves.}

\subsection{Critical points and tracker conditions}
Critical points are obtained by simultaneously solving the equations in~(\ref{eq:pol}) when all derivatives of the variables on the left side are zero. From equations~\eqref{eq:pol_a} and~\eqref{eq:pol_d}, one can see from $\theta_{\phi}^\prime = \theta_{\sigma}^\prime = 0$, that the critical conditions $y_{1 \phi c} = 3 \sin{\theta_{\phi_c}} $ and $y_{1 \sigma c} = - 3 \sinh{\theta_{\sigma_c}}$ are common to all fixed points. Using the forms of $y_{2\phi}$ and $y_{2\sigma}$ given in equations~\eqref{eq:GP}, one can derive from $y_{1\phi}^\prime=y_{1\sigma}^\prime=0$ the following two equations;

\begin{align} \label{eq:trach_quint}
&\left(\gamma_{t o t}+\frac{\alpha_0}{9} \Omega_{\phi c}  \right. \nonumber \\ 
  &\left. \quad +\frac{2}{3} \alpha_1 \Omega_{\phi c}^{1 / 2} \sin(\theta_{c}/2) +4 \alpha_2 \sin^2 (\theta_{c}/2)\right) \sin \theta_c=0 \, ,
\end{align}
and
\begin{align} \label{eq:tracck_phan}
&\left(\gamma_{t o t}-\frac{\beta_0}{9} \Omega_{\sigma c} \right. \nonumber \\
  &\left. \quad +\frac{2}{3} \beta_1 \Omega_{\sigma c}^{1 / 2} \sinh \left(\theta_c / 2\right)-4 \beta_2 \sinh ^2\left(\theta_c / 2\right) \right) \sinh \theta_{\sigma c}=0  \, .
\end{align}

In general, critical points can be classified into three different classes depending on the dominant energy densities. The fluid dominated fixed points ($\Omega_{\phi c} = \Omega_{\sigma c} =0$), the quintessence ($\Omega_{\phi c}=1$), and the phantom field ($\Omega_{\phi c}=1$) dominated fixed points. In Appendix \ref{appen:fixed_eigen} we have listed all possible fixed points of the system and discussed the stability of these fixed points. Here, we are going to discuss in detail the fluid-dominated fixed points as they correspond to the tracking behavior of the scalar fields. 

It has been shown in~\cite{Urena-Lopez:2020npg,LinaresCedeno:2021aqk} that Eqs.~\eqref{eq:trach_quint} and~\eqref{eq:tracck_phan}, for the quintessence and phantom models, respectively, can give us a generalized tracking behavior for any choice of the parameters $\alpha$ and $\beta$ once we consider $\Omega_\phi \simeq \Omega_\sigma \simeq 0$ in the early universe. However, the equivalent simplest tracker condition can be calculated considering $\alpha_0 = \alpha_1 = 0$, $\beta_0 =\beta_1 = 0$ and using them in equations~\eqref{eq:trach_quint} and~\eqref{eq:tracck_phan}. For the quintessence field and the phantom field, the tracker condition will be, respectively,
\begin{equation}
    \gamma_{\phi c} = - \gamma_{tot}/(2 \alpha_2) \, , \quad \gamma_{\sigma c} = - \gamma_{tot}/(2 \beta_2) \, . \label{eq:trackerEoS}
\end{equation}

For the case of the quintessence field, an upper bound can be placed on the parameter $\alpha_2$ by considering the fact $0<\gamma_{\phi} <2$ and this will translate into $\alpha_2 < -\gamma_{tot} / 4$. For the tracking to start at the early radiation domination $\gamma_{tot} = 4/3$ and $\alpha_2 < -1/3$. For the phantom field, the condition would simply be $\beta_2 >0$. 

%A more general equivalent condition for tracking can be obtained . This condition can be easily achieved from the critical condition applied to Eqs.~\eqref{eq:pol_c} and~\eqref{eq:pol_f},
%\begin{subequations}
%\begin{eqnarray}
%    (\gamma_{tot} - \gamma_{\phi c}) \Omega_{\phi c} =0 \, , \\
%    (\gamma_{tot} - \gamma_{\sigma c}) \Omega_{\sigma c} =0 \, .
%\end{eqnarray}
%\end{subequations}
%The critical points that correspond to $\Omega_{\phi c} = \Omega_{\sigma c} = 0$ are the fixed points dominated by the %fluid component, and this allows the fields to track the background evolution during the eras of radiation and matter %domination. 

\subsection{Fluid-dominated critical points} \label{sec:track}
In Table~\ref{tab:matter_fixed} we have listed all fixed points dominated by background fluids $m_1$ to $m_4$. The fluid-dominated fixed points are characterized by $\Omega_{\phi c} = \Omega_{\sigma c} =0$, which means that the universe is dominated by the background fluids in their corresponding eras, such as radiation domination, matter domination, and the contribution to the total energy density of the quintessence and phantom fields is negligible. During radiation domination $\gamma_{tot} = 4/3$ and matter domination $\gamma_{tot} = 1$. One can see from Table~\ref{tab:matter_fixed} that there are four matter-dominated fixed points corresponding to four physically interesting scenarios depending on the tracking behavior of the scalar fields.

\begin{table}[]
\caption{ List of fluid dominated fixed points. For fixed point $m_1$, no tracker behaviour of the scalar field is obtained, for $m_2$ (phantom) and and $m_3$ (quintessence) only one field shows tracker behaviour. For $m_4$ fixed point both the scalar fields shows tracker behaviour.}
\label{tab:matter_fixed}
\begin{tabular}{|c|c|c|c|c|}
\hline
\multicolumn{1}{|l|}{\begin{tabular}[c]{@{}l@{}}Fixed \\ Points\end{tabular}} & $\Omega_{\phi_c}$ & $\sin \theta_{\phi_c}$ & $\Omega_{\sigma_c}$ & $\sinh \theta_{\sigma_c}$ \\ \hline
$m_1$ & 0 & $n \pi$ & 0 & 0 \\ \hline
$m_2$ & 0 & $n \pi$ & 0 & $\sinh^2(\theta_{\sigma_c}  / 2) = \frac{\gamma_{tot}}{4 \beta_2}$ \\ \hline
$m_3$ & 0 & $\sin^2(\theta_{\phi_c}  / 2) =- \frac{\gamma_{tot}}{4 \alpha_2}$ & 0 & 0 \\ \hline
$m_4$ & 0 & $\sin^2(\theta_{\phi_c}  / 2) =- \frac{\gamma_{tot}}{4 \alpha_2}$ & 0 & $\sinh^2(\theta_{\sigma_c}  / 2) = \frac{\gamma_{tot}}{4 \beta_2}$ \\ \hline
\end{tabular}
\end{table}

\begin{figure*}[htb]

\subfloat[\label{fig:no_track}]{%
  \includegraphics[width=\columnwidth]{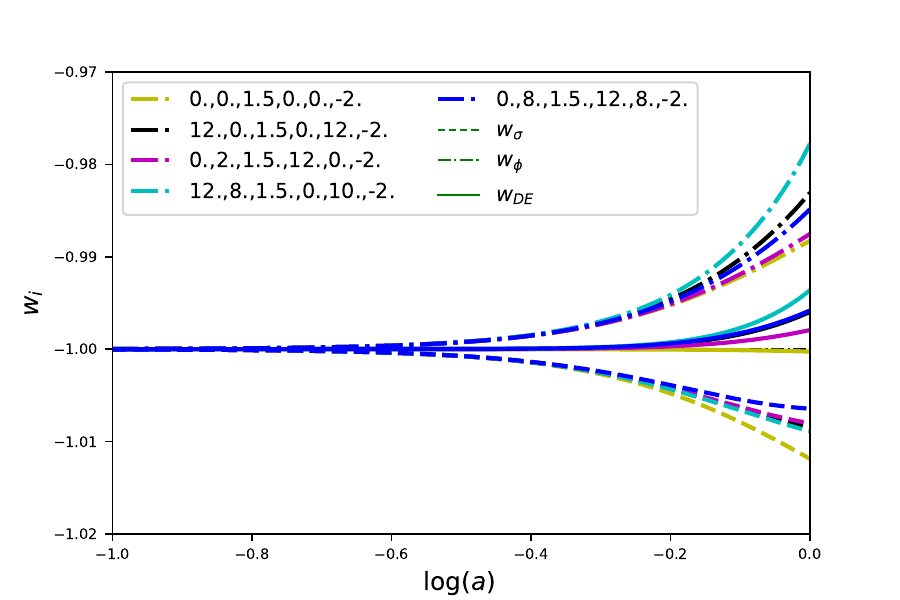}%
}\hfill
\subfloat[\label{fig:phan_track}]{%
  \includegraphics[width=\columnwidth]{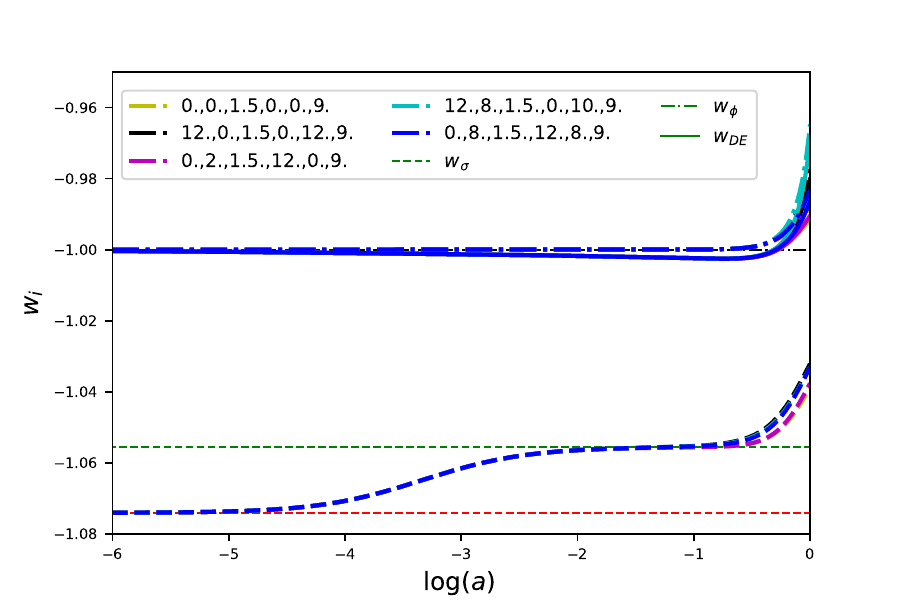}%
}

\subfloat[\label{fig:quint_track}]{%
  \includegraphics[width=\columnwidth]{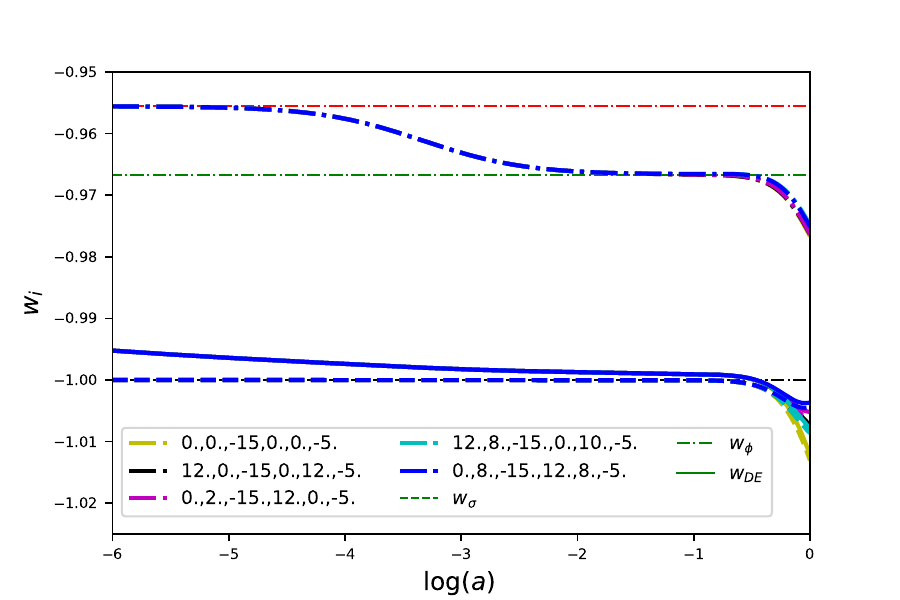}%
}\hfill
\subfloat[\label{fig:both_track}]{%
  \includegraphics[width=\columnwidth]{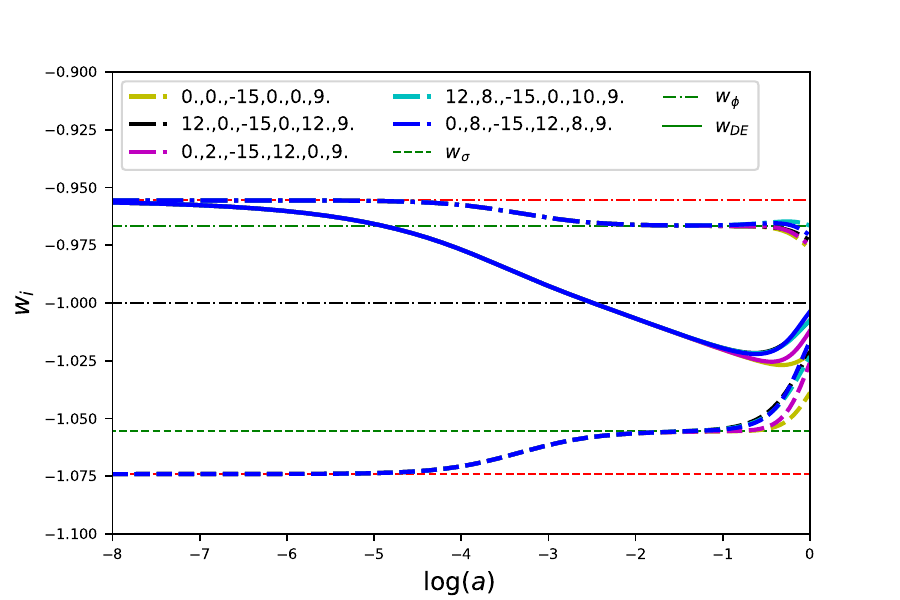}%
}

\caption{ The following plots show the equations of state (EoS) for the total dark energy and its component scalar fields (quintessence and phantom) under different choices of $(\alpha, \beta)$ parameters. In (a), the solution that originated at the fixed point $m_1$ is displayed where the scalar field does not show the tracking behavior. In (b), the solutions from the fixed point $m_2$ are shown where only the phantom field shows tracking behavior. In (c), the solutions from the fixed point $m_3$ are shown where only the quintessence field tracks. Finally, in (d), the solutions from the $m_4$ fixed point are shown, in which both fields track the background. }\label{fig:track}
\end{figure*}

The fixed point $m_1$ corresponds to a scenario where the quintessence and phantom fields behave like a cosmological constant and there is no tracking phenomenon from any fluid.  In Fig.~\ref{fig:no_track} we have plotted the EoS of the total dark energy and its components for the solutions that originate from the fixed point $m_1$ for various choices of the parameters $\alpha$ and $\beta$. Here, we make sure that $\alpha_2 > -1/3$ and $\beta_2 <0$ to turn off the tracker behavior. As expected from the analytical point of view, there is no tracker behavior and the total dark energy EoS $w_{DE}$ remains indistinguishable from the cosmological constant, but at a late time it starts to evolve and currently $w_{DE}>-1$. 

At $m_2$ the dominating fluid is tracked by the phantom field, but the quintessence field remains indistinguishable from the cosmological constant in both the radiation and matter dominated eras. In Fig.~\ref{fig:phan_track}, we show that the solutions originate at the critical point $m_2$ for different parameters $\alpha, \beta $ with the choice of $\alpha_2 > -1/3$ and $\beta_2 > 0$. The expected behavior described above in the critical point analysis can be seen in the evolution of the numerical solutions. For this case, currently $w_{DE} > -1$

Similarly, the $m_3$ quintessence field tracks the fluid counterpart, while the phantom field remains indistinguishable from the cosmological constant during radiation and matter domination, but it shows deviation at the late time. In Fig.~\ref{fig:quint_track}, we have plotted the dynamics of the EoS of the quintom field together with the components for different values of the parameters $\alpha, \beta $, with $\alpha_2 < -1/3$ and $\beta_2 < 0$, so that the quintessence field satisfies the tracker condition while the phantom field does not. 

The most interesting case is the solutions originating from the fixed point $m_4$, in which both the quintessence and the phantom fields track the background by satisfying the conditions $\alpha_2 < -1/3$ and $\beta_2 > 0$. In Fig.~\ref{fig:both_track}, we have plotted the EoS of both the scalar fields and the total dark energy. It can be seen that both quintessence and phantom fields track the background, and the corresponding EoS is decided by the tracker condition they follow. The total dark energy EoS shows interesting behavior; deep in the radiation-dominated era, it tracks the evolution of the quintessence field followed by a phantom barrier crossing and the current value of $w_{DE} < -1$.

\begin{figure}[h] 
%\centering
\includegraphics[width=\columnwidth]{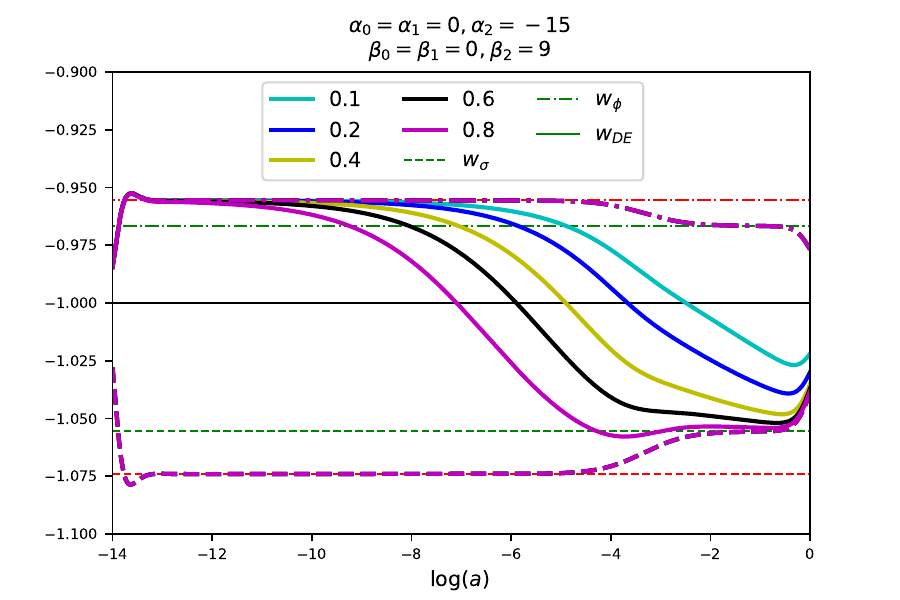}
\caption{\label{fig:fplot} Plots of EoS of the quintessence field, the phantom field and the total DE EoS for different choices of the $f$ parameter when both the fields are tracking the background. See the text for more details.}
\end{figure}

It is interesting to note here that the fixed points $m_1$ to $m_3$ are reported in the previous section while discussing the fixed points of quintessence and phantom domination separately and also in \cite{Urena-Lopez:2020npg,LinaresCedeno:2021aqk}. The fixed point $m_4$ appears when both the phantom and quintom fields are switched on in the evolution of the universe. The choice of parameters $\alpha_2$ and $\beta_2$, for all the figures mentioned in this section, is also within the posterior range obtained from the statistical analysis in the next section. 

\subsection{Initial conditions}
Here, we discuss the method for finding the initial condition of the system by solving Eqs.~(\ref{eq:pol}) using the tracker condition. An estimation of the proper initial conditions is necessary for the implementation of the model in the Boltzmann code called the Cosmic Linear Anisotropy Solving System (\textsc{CLASS})~\cite{lesgourgues2011cosmic,Lesgourgues:2011re,Lesgourgues:2011rg,Lesgourgues:2011rh}. 

During the era of matter and radiation domination, both quintessence and the phantom field contribute negligibly to the energy density, and the total EoS remains constant at $\gamma_{tot} =4/3$ for radiation and $\gamma_{tot}=1$ for matter domination. We can use this fact and the tracking condition for the quintessence and phantom fields (see Eqs.~\eqref{eq:trackerEoS}) to estimate the initial conditions for which a viable evolution can be achieved. We have used the scheme mentioned above to approximate the solution of the equations in Eq.~\eqref{eq:pol} at the radiation and matter domination, and match them at the radiation matter equality to estimate the initial conditions. The initial condition for the quintessence field would be the following;
\begin{subequations}
\label{eq:ini}
\begin{equation}
\begin{aligned}
\cos \theta_i &=1+\frac{2}{3 \alpha_2} \quad y_{1 i}=3 \sin \theta_i \\
\Omega_{\phi i} &= a_i^{4\left(1+1 / 2 \alpha_2\right)}\left(\frac{\Omega_{m 0}}{\Omega_{r 0}}\right)^{1+1 / 2 \alpha_2} \Omega_{\phi_0}
\end{aligned} \label{eq:quint_ini}
\end{equation}
For the phantom field, the initial conditions are,
\begin{equation}
\begin{aligned}
\cosh \theta_i &=1+\frac{2}{3 \beta_2}, \quad y_{1 i}=-3 \sinh \theta_i \\
\Omega_{\phi i} &=a_i^{4\left(1+1 / 2 \beta_2\right)}\left(\frac{\Omega_{m 0}}{\Omega_{r 0}}\right)^{1+1 / 2 \beta_2} \Omega_{\sigma_0}
\end{aligned} \label{eq:phantom_ini}
\end{equation}
\end{subequations}
where $\Omega_{\phi 0} = A_\phi (1-f) \Omega_{DE 0} $ and $\Omega_{\sigma 0} = A_\sigma f \Omega_{DE 0}$. In addition, $f$ is an extra sampling parameter which we introduce to have better control over the contributions of $\Omega_{\phi i}$ and $\Omega_{\sigma i}$ to the total density.

% My suggestion is that we use one of the entries in the vector scf\_parameters[] as an extra sampling parameter. Let us call it $f$ so that $0 < f < 1$, then my proposal is . The prior for the extra parameter would be something like $f=[0.1:0.9]$ or even $f=[0.01:0.99]$. If the result is $f \sim 1$ ($f \sim 0$) then the dominant component is phantom (quintessence). Another advantage is that the total EoS currently would be given by $\gamma_{DE} = \gamma_\phi (1-f) + \gamma_\sigma f$, and again $f$ would signal any preference for quintessence or phantom.

The background equations of motion~\eqref{eq:pol}, together with the initial conditions~\eqref{eq:ini}, were incorporated into an amended version of the code \textsc{CLASS} following the prescription already described in previous work, for example~\cite{Urena-Lopez:2020npg,LinaresCedeno:2021aqk,roy2018new}. The present values of the matter density and radiation densities are represented by $\Omega_{r0}$ and $\Omega_{m0}$, respectively. 

One difference with respect to single-scalar field models is that we need to adjust the initial conditions of the quintom fields, so that the solution of the equations of motion delivers the right present value of the total DE density parameter; i.e. $\Omega_{DE 0}$, as the sum of the present value of the quintessence energy density $\Omega_{\phi 0}$ and the phantom energy density $\Omega_{\sigma 0}$. For that, we let the internal algorithm of \textsc{CLASS} find the appropriate value of the coefficients $A_\phi$ and $A_\sigma$ for a given choice of the parameter $f$. The initial value of the scale factor is typically considered $a_i \simeq 10^{-14}$ in the \textsc{CLASS} code.

% where we have assumed $\Omega_{\phi 0} = A_\phi \Omega_{DE 0} $ and $\Omega_{\sigma 0} = A_\sigma \Omega_{DE 0}$.  

In Figure~\ref{fig:fplot}, we have displayed the development of the EOS of both the scalar field and the total dark energy EOS for different values of the $f$ parameter. It is evident that the individual EOS of the fields is not affected by different choices of $f$, however, the total dark energy EOS is impacted and the epoch of phantom barrier crossing is sensitive to it. This is because the definition of the $w_{tot}$ (see Eq.~(\ref{eq:wtot})) involves the $\Omega_\phi$ and $\Omega_\sigma$. A lower value of $f$, which corresponds to $\Omega_{\phi i} >> \Omega_{\sigma i}$, causes the crossing of the phantom barrier to be shifted further towards the matter-dominated era.

\section{Constraints on cosmological parameters \label{sec:constraints}}
We have used current cosmological observations to constrain the quintom model presented here by using an amended version of the Boltzmann code \textsc{class} and the cosmological MCMC sampler \textsc{MontePython} (v3.5)~\cite{Audren:2012wb}. The cosmological observations that have been used for this are PantheonPlus~\cite{Brout:2022vxf} together with SH0ES prior\cite{Riess:2021jrx}, BAO (BOSS DR12~\cite{Alam_2017},  eBOSS DR14 (Lya)~\cite{Cuceu_2019}) and the WiggleZ galaxy survey \cite{Kazin_2014}. We have also used a compressed Planck likelihood by following the proposal in~\cite{Arendse_2020} (see their Appendix~A). The parameter constraining accuracy of compressed Planck likelihood is the same as the full Planck data, but consumes less time and computational resources.

We have considered flat priors for the cosmological parameters $100~\omega_{b}:[1.9,2.5], \omega_{cdm}:[0.095,0.145]$. Following the suggestion in~\cite{Sabti:2021xvh}, we fix the angular scale of the sound horizon $\theta_{s}$ to the Planck CMB value of $\theta_{s} = 1.04110$ \cite{Aghanim:2018eyx}, and the current value of the Hubble parameter $H_0$ as the derived parameter. Since $\theta_{s}$ is determined by the angular scales of the acoustic peaks in the temperature and polarization spectra as a purely geometrical quantity, and its determination is mostly independent of the underlying physics considered for the CMB era, fixing it does not make any significant changes to the results.

A flat prior of $[-20, 20]$ has been considered for the parameters $\alpha_2$ and $\beta_2$. Since it has already been shown in~\cite{roy2018new,roy2019arbitrariness,Urena-Lopez:2019xri,LinaresCedeno:2020dte} that the parameters $\alpha$ and $\beta$ remain unconstrained for quintessence and phantom, we consider $\alpha_0 = \alpha_1 =0$ and $\beta_0=\beta_1=0$. Parameters $\alpha_2$ and $\beta_2$ are the most interesting, as they determine the tracking behavior of the scalar fields. One can also see from our choice of priors that it includes all four tracking scenarios described in the previous section.

The constraints obtained from the observations are given in Table~\ref{tab:param}, with their mean value and the 68\%CL. For parameter estimation, we have used the publicly available MCMC analyzer code \textsc{GetDist}\cite{Lewis:2019xzd}.

\begin{table}[]
\caption{ 68\%CL constraints on the cosmological parameters for the $\Lambda$CDM and the quintom models using the combining data sets of Pantheon Plus Survey~\cite{Brout:2022vxf} with SH0ES prior\cite{Riess:2021jrx}, Baryon Acoustic Oscillations from the BOSS DR12 survey ~\cite{Alam_2017}, Lyman-alpha forest data from the eBOSS DR14 survey ~\cite{Cuceu_2019}, and the WiggleZ galaxy survey \cite{Kazin_2014}, along with the compressed Planck likelihood.\label{tab:param}}

\begin{tabular} { |l| c| c|}
\hline
 Parameter & $\Lambda$CDM & Quintom\\
\hline
{\boldmath$10^{-2}\omega{}_{b }$} & $2.260\pm 0.015            $&$2.250^{+0.015}_{-0.013}   $\\
{\boldmath$\omega_{cdm}          $} & $0.11719^{+0.00086}_{-0.00077}$ & 
$0.11832^{+0.00091}_{-0.0011}$\\
{\boldmath$H_0            $} & $68.35^{+0.33}_{-0.37}     $&$69.56^{+0.21}_{-0.96}     $\\

{\boldmath$\Omega_{DE}    $} &  $0.6992\pm 0.0064          $&$0.7070^{+0.0040}_{-0.0082}$\\

{\boldmath$\Omega_{\phi}  $} & --&$0.0966^{-0.0055}_{-0.14}  $\\

{\boldmath$\Omega_{\sigma}$} & --&$0.61^{+0.12}_{-0.39}      $\\

{\boldmath$w_{DE}         $} & -1&$-1.047^{+0.035}_{-0.0098} $\\

{\boldmath$w_{\phi}       $} & --&$-0.9536^{+0.0061}_{-0.064}$\\

{\boldmath$w_{\sigma}     $} & --&$-1.058^{+0.044}_{-0.0076} $\\

{\boldmath${\alpha_2}     $} & --& $-2^{+12}_{-17}            $\\

{\boldmath${\beta_2}      $} & --& $9.1^{+3.5}_{-5.8}         $\\

{\boldmath$\delta{\chi_{min} ^2}     $} & 0 &$-6.0  $\\

{\boldmath$\ln{B_{Q\Lambda}}     $} & - &$1.21 $\\
\hline
\end{tabular}
\end{table}

\begin{figure}[t] 
%\centering
\includegraphics[width= 0.9\columnwidth]{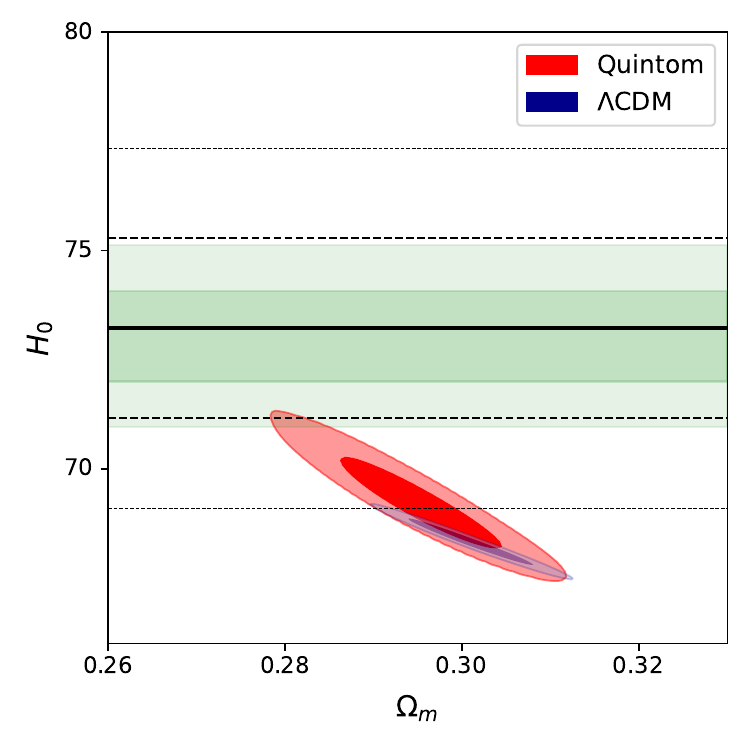}
\caption{\label{fig:hubble} Two dimensional posterior plot of the $H_0$ versus $\Omega_m$ parameters. The horizontal gray  band shows the constraint on the $H_0$ parameter from the SH0ES collaboration~\cite{Riess:2019cxk}. See the text for more details.}
\end{figure}

In Fig.~\ref{fig:hubble}, we show the contour plot of the posterior of Hubble parameter $H_0$ and $\Omega_m$, with the quintom model in red and the $\Lambda$CDM model in blue. The green band shows the $1\sigma$ (deep green) and $2\sigma$ (light green) constraints on $H_0$ from~\cite{Riess:2021jrx}, whereas the region covered by the black dashed lines shows the $1\sigma$ and $2 \sigma$ constraints obtained from~\cite{Scolnic:2023mrv}.  There is a significant shift of the current value of the Hubble parameter towards the higher value. We use the following estimator of the tension given in~\cite{Camarena:2018nbr} to quantify the $H_0$ tension;

\begin{equation}
T_{H 0}=\frac{\left|H_0-H_0^{loc}\right|}{\sqrt{\sigma_{H 0}^2+\sigma_{\mathrm{loc}}^2}}, \nonumber
\end{equation}
where $T_{H0}$ is the amount of tension and $H_0$ is the mean and $\sigma_{H 0}^2$ is the variance of the posterior $p(H_0)$. 

Once we compare our results with the current value of the Hubble parameter ($H_0=73.04 \pm 1.04 \mathrm{~km} \mathrm{~s}^{-1} \mathrm{Mpc}^{-1}$) reported in~\cite{Riess:2021jrx}, from the Hubble Space Telescope and the SH0ES data, the tension decreases to $T_{H0} \simeq 2.85 \sigma$. Compared to the Hubble value ($H_0=73.22 \pm 2.06 \mathrm{~km} \mathrm{~s}^{-1} \mathrm{Mpc}^{-1}$) reported in~\cite{Scolnic:2023mrv}, from the standardized TRGB and Type Ia supernova data sets, the tension reduces to $T_{H0} \simeq 1.67 \sigma$. According to Table IV in~\cite{Camarena:2018nbr} for the first case, the tension is moderate in nature and for the second case the tension is weak. 

\begin{figure*}[t] 
%\centering
\includegraphics[width= 2\columnwidth]{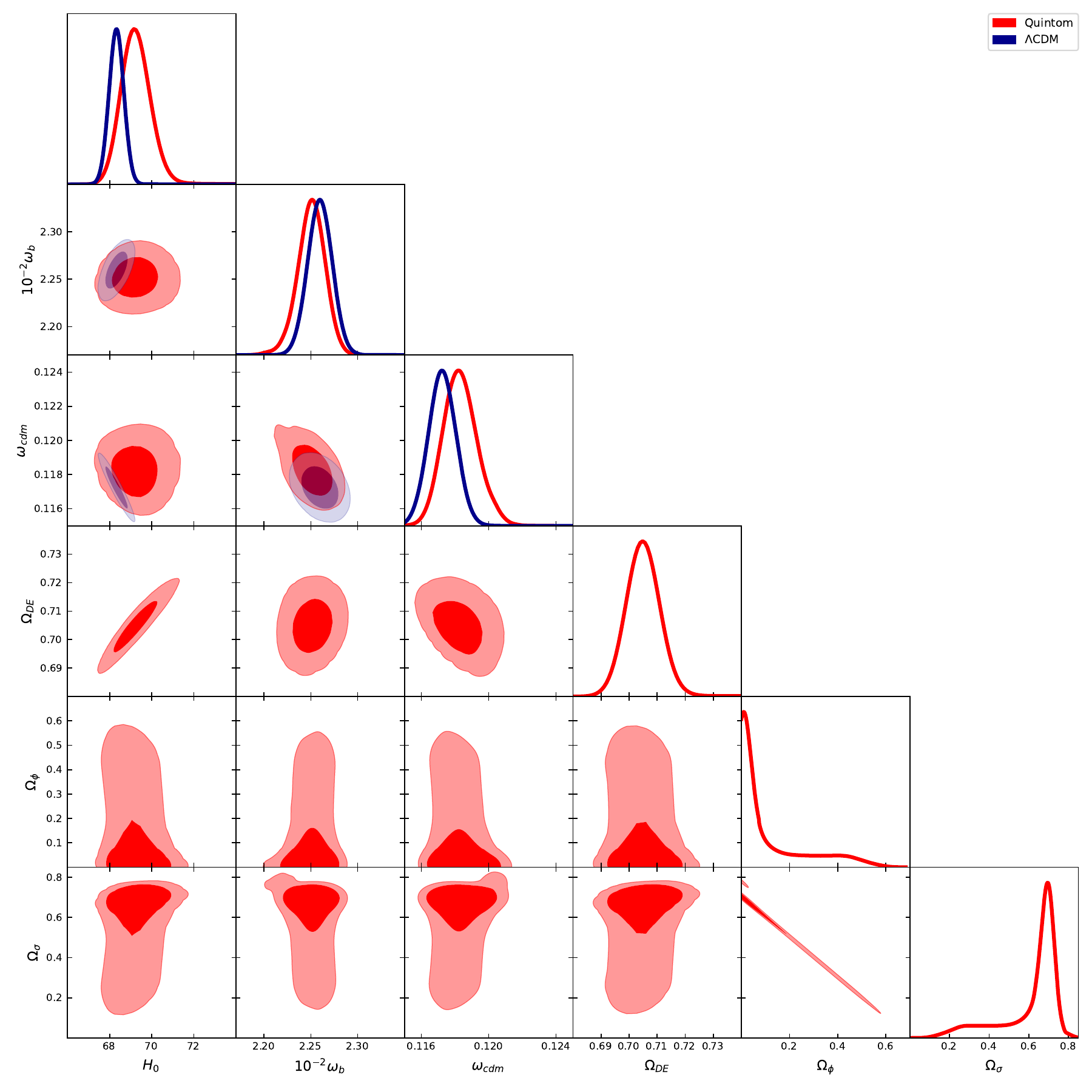}
\caption{\label{fig:cosmo} Triangular plot displays 1D and 2D posteriors of the cosmological parameters using Pantheon~\cite{Scolnic:2017caz}, BAO (DR12~\cite{Alam_2017}, DR14~\cite{Cuceu_2019}), WiggleZ~\cite{Kazin_2014} considering SH0ES and compressed Planck prior. The blue contours and curves are for $\Lambda$CDM and the red ones are for the quintom model.}
\end{figure*}

\begin{figure}[h] 
%\centering
\includegraphics[width= 
\columnwidth]{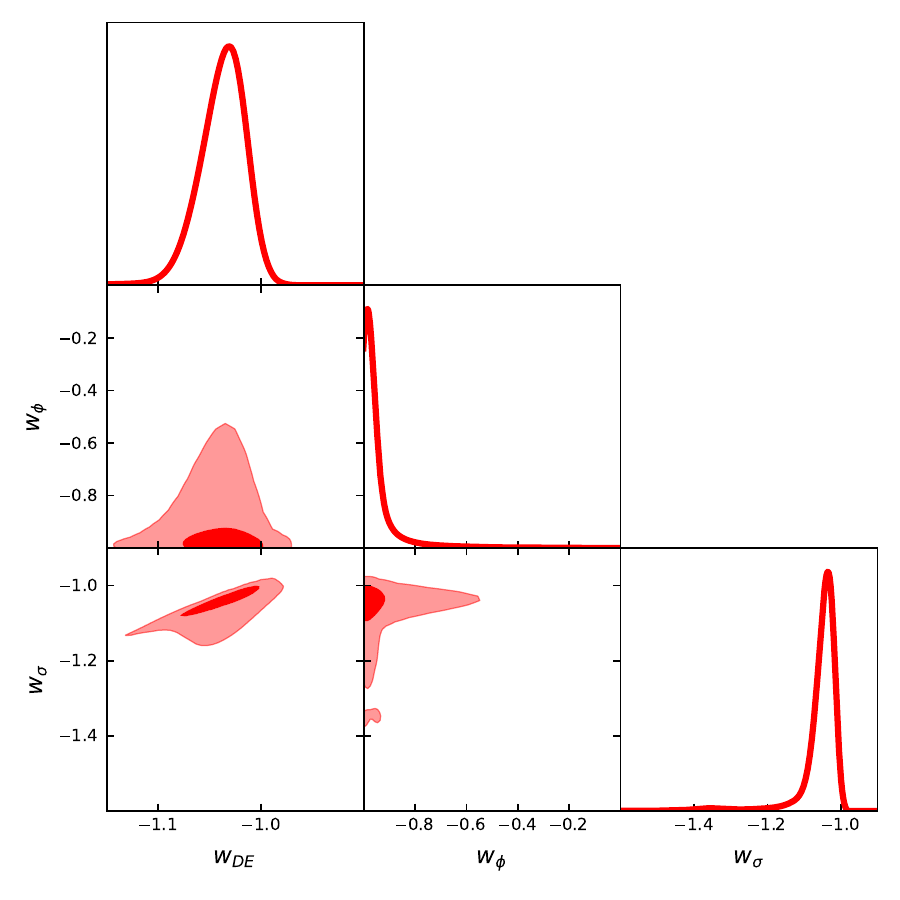}
\caption{\label{fig:model} Triangle plot showing 1D and 2D posteriors of the EoS of the total dark energy($w_{DE}$) and its components $w_\phi$ and $w_\sigma$. }
\end{figure}

The triangular plots in Fig.~\ref{fig:cosmo} show the 1D and 2D posteriors of the cosmological parameters $H_0, 10^{-2} w_b, w_{cdm},\Omega_{DE}, \Omega_{\phi}, \Omega_{\sigma}$. A comparison with the standard $\Lambda$CDM case has been shown by plotting the constraints on the parameters in blue contours. In Fig.~\ref{fig:model}, we have shown the posterior distribution of the total dark energy EoS $w_{DE}$, the quintessence EoS $w_\phi$, the phantom EoS $w_\sigma$, and of the parameter $f$. 

\begin{figure}[ht] 
%\centering
\includegraphics[width=\columnwidth]{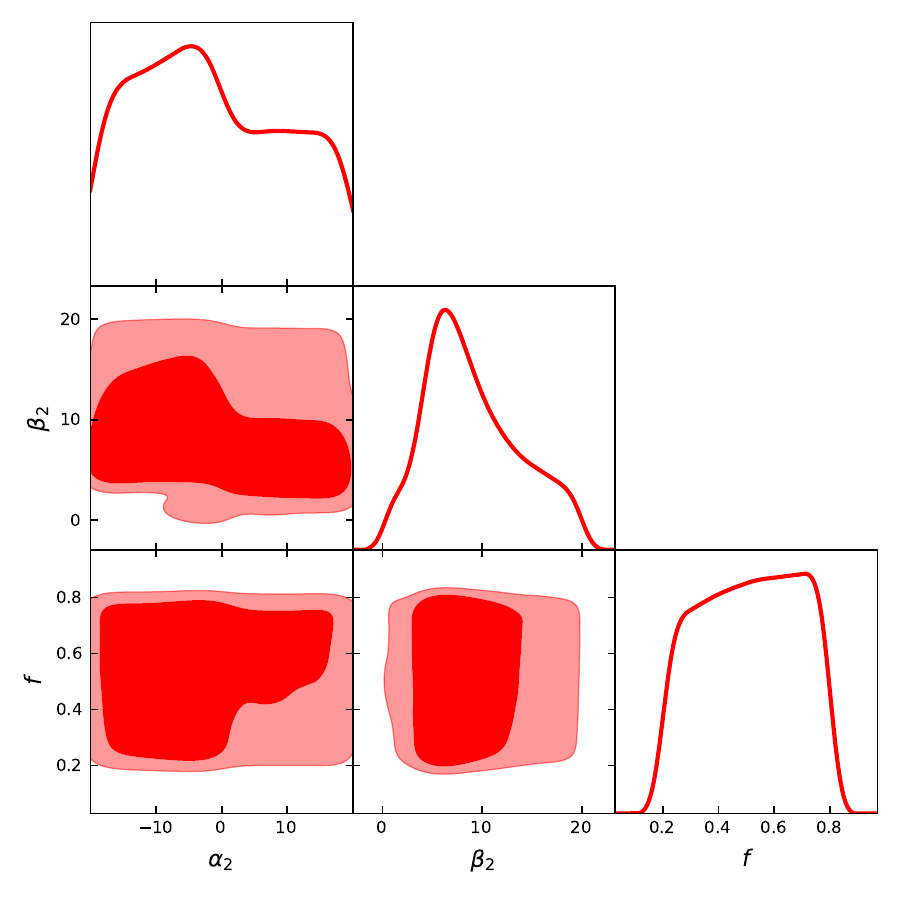}
\caption{\label{fig:alpha} Triangle plot showing 1D and 2D posteriors of the two tracking parameters $\alpha_2$ for the quintessence and $\beta_2$ for the phantom scalar field. Notice the unconstrained nature of parameters $f$ and $\alpha_2$, and the existence of a preferred value for $\beta_2$. See the text for more details.  }
\end{figure}

The value of $w_{DE}$ remains in the phantom region (see Table~\ref{tab:param}). An interesting fact to be observed here from Fig.~\ref{fig:alpha} is that $\alpha_2$ remains unconstrained, whereas $\beta_2 \simeq 9.1^{+9.9}_{-7.8} \, (2\sigma)$ indicates the tracking behavior of the phantom field. The unconstrained $\alpha_2$ can be explained by the fact that the current observations require the EoS of the dark energy to be $w_{DE} < -1$ and the quintessence field should not dominate the late-time dynamics. 

Though the phantom field dominates late-time cosmology, the contribution of the quintessence field to the current budget is non-negligible, as it is approximately $15\%$ of the total dark energy budget. There could be two physical scenarios related to the unconstrained nature of $\alpha_2$. The quintessence field tracks the background during radiation domination, as mentioned in Sec.~\ref{sec:track} and Fig.~\ref{fig:both_track}. This case will always be associated with a phantom barrier crossing. In the other scenario, the quintessence field does not track the background, the total $w_{DE}$ remains indistinguishable from the cosmological constant case $w_{DE} = -1$, and at the late time becomes phantom in nature (Sec.~\ref{sec:track} and Fig.~\ref{fig:phan_track}). 

As can be seen by comparing Figs.~\ref{fig:quint_track} and~\ref{fig:phan_track}, when the phantom field tracks the background but the quintessence field does not, the current value of the EoS satisfies $w_{DE} > -1$. Since observational constraints suggest the current value of $w_{DE} < -1$, it is more reasonable for both fields to track the background, even though observations cannot constrain $\alpha_2$. Due to the lower bound of $\beta_2>1.3$, it can be checked that the epoch of the crossing of the phantom barrier should occur after the radiation-matter equality. In Fig.~\ref{fig:alpha}, we have also shown the posterior distribution of the parameter $f$, which is related to the distribution of energy between the phantom and the quintessence at the initial stage. Although the dynamics of the scalar fields are sensitive to the $f$ parameter, $f$ remains unconstrained by observations. The possible reason behind this might be that for any particular choice of the $f$ there are some values of $\alpha_2, \beta_2$ that help the quintom model to satisfy the observational constraints.

 We have used $\chi^2_{min}$ (see Table~\ref{tab:param}) to compare the $\Lambda$CDM model with the quintom model. The difference in value between the models, $\Delta \chi^2_{min} = \chi^2_{Quintom} - \chi^2_{\Lambda CDM} = -6$, suggests a better fit to the data by the quintom model compared to $\Lambda$CDM. We also computed the Bayes factor to be certain of the preference for the model chosen by the data, so that the presence of extra model parameters for the quintom model does not affect it. We have calculated $\ln B_{Q \Lambda} = \ln \mathcal{Z}_{Q} - \ln \mathcal{Z}_\Lambda$, where $\mathcal{Z}$ is Bayesian evidence, and the suffix $Q$ and $\Lambda$ represent the quintom and $\Lambda$CDM models, respectively. 
 
 In general, Jeffrey's scale is used to find the preference of one model over another. The strength of preference of the quintom model over $\Lambda$CDM will depend on the value of $\ln B_{Q \Lambda}$.  The preference is negative if $\ln B_{Q \Lambda} <1$, and if $\ln B_{Q \Lambda}>1; >2.5; >5.0$ the preference is positive, moderate, and strong, respectively. For more details on Bayesian model selection, see~\cite{Trotta:2005ar}.  Since the calculation of the Bayes factor is computationally challenging, we have used the publicly available code \textsc{MCEvidence}~\cite{Heavens:2017afc}. This code can be used directly to calculate the Bayes factor from the MCMC chains generated by \textsc{MontePython}. In the comparison of Quintom with $\Lambda$ CDM, we find $\ln B_{Q \Lambda} = 1.21$. This shows that the quintom model would be positively favored by the observations compared to $\Lambda$CDM.

%In Fig.~\ref{fig:Hz}, we can see the evolution of the composite quantity $H(z)/(1+z)$ as a function of the redshift $z$, for the set of parameters $\alpha_2$ and $\beta_2$ that are within the posterior obtained from MCMC analysis. To compare with observational data, we have plotted the results obtained from the SH0ES~\cite{Riess:2019cxk} and BAO observations~\cite{BOSS:2016wmc,Zarrouk:2018vwy,Blomqvist:2019rah,deSainteAgathe:2019voe}. The blue dashed line corresponds to the $\Lambda$CDM model, whereas the solid lines correspond to the quintom models.

%\begin{figure}[h] 
%\centering
%\includegraphics[width= \columnwidth]{Hz.pdf}
%\caption{\label{fig:Hz} Plot of $H(z)/(1+z)$ against $z$ for the %set of parameters $\alpha_2$ and $\beta_2$ chosen within the %posteriors obtained from the data analysis.  }
%\end{figure}
 
\section{Conclusions \label{sec:conclusions}} 
In this work, we have studied the dynamics of the quintom scalar field model and used the current cosmological observations to constrain the cosmological parameters. Quintom dark energy model consists of two scalar fields; one is the canonical scalar field called quintessence and the other is the phantom scalar field. For both of these scalar fields, the choice of potentials remains arbitrary, since there is no observational preference for any particular form of the potential~\cite{roy2018new,LinaresCedeno:2021aqk,roy2019arbitrariness}, and we have then considered the general parameterization of potentials proposed in~\cite{roy2018new}. The set of equations has been converted into an autonomous system using suitable transformations of the variables. For the quintessence sector, a polar transformation has been considered, while a hyperbolic one was required to incorporate the negative kinetic energy of the phantom field. 

We have discussed the fixed points of the model and the general phase-space behavior, in particular, in finding the existence of the tracker solutions for the quintom model. There could be four physically interesting scenarios: no tracking behavior of either field, either one field tracks the background while another does not, or both fields track the background. We have also found the tracking conditions for quintessence and phantom fields separately, which has previously been reported in the literature. Of the four cases, the most interesting is the case in which both fields track the background, with the quintessence field contributing more to the dark energy density at early times, whereas the phanton field dominates the universe currently. If both fields track the background, there should be a phantom barrier crossing of the total dark energy EoS which is most likely to happen after the matter-radiation equality era.

The available cosmological observations have been used to constrain the cosmological parameters. Our choice of priors includes both the tracking and no-tracking behavior of the quintom and the phantom field. The observations suggest that at late times the phantom field should dominate the dark energy sector with an approximately $15\%$ share to the quintessence counterpart, and with both fields tracking the background at early times. 

We have also reported the current status of the Hubble tension in the quintom model. Although it is not possible to completely resolve the Hubble tension in this model, it can be reduced to the level of $2.6 \sigma$ compared to the Hubble value reported in~\cite{Riess:2021jrx}, and to $1.6 \sigma$ when compared with~\cite{Scolnic:2023mrv}, making it a moderate and week tension, respectively, comparable to the strong tension of $\Lambda$ CDM. We have also studied the selection of Bayesian models by computing the Bayes factor to compare with $\Lambda$CDM and found a positive preference for the quintom model. 

For completeness, a thorough dynamical system analysis of the model is included in the Appendix. It is possible that the late-time attractors are either quintessence- or phantom-field dominated. Since current observations indicate that the universe is currently phantom dominated, the phantom attractor points are more pertinent. It would be beneficial to incorporate new data sets and determine whether the preference for phantom domination continues, as well as to gain a better understanding of the active parameters of our model. This research will be conducted in the future and reported elsewhere.

\begin{acknowledgments}
We acknowledge the use of the Chalawan High-Performance Computing cluster, operated and maintained by the National Astronomical Research Institute of Thailand (NARIT). The research of LAU-L was partially supported by Programa para el Desarrollo Profesional Docente; Dirección de Apoyo a la Investigación y al Posgrado, Universidad de Guanajuato; CONACyT México under Grants No. A1-S-17899, No. 286897, No. 297771, No. 304001; and the Instituto Avanzado de Cosmología Collaboration.
\end{acknowledgments}

%\newpage
\appendix

%\section{Energy density for multi-scalar fields}
%Here we discuss the way to express the density of a particular scalar field in a multi-scalar field model when the $\Omega_{\phi i}$ is known. This is the approach in which we have implemented the CLASS code to calculate the energy density of quintessence and phantom fields. 

%We start by writing the definition of the density parameters for a set of $M$ scalar fields as
%\begin{eqnarray}
%  \rho_1 &=& \Omega_1 \left( \rho_r + \rho_m + \rho_\Lambda + %\sum^M_{j=1} \rho_j \right) \, , \\
%  \rho_2 &=& \Omega_2 \left( \rho_r + \rho_m + \rho_\Lambda + %\sum^M_{j=1} \rho_j \right) \, , \\
%  && \ldots \, , \\
%  \rho_M &=& \Omega_M \left( \rho_r + \rho_m + \rho_\Lambda + %\sum^M_{j=1} \rho_j \right) \, .
%\end{eqnarray}
%The above system of equations can be written as a set of linear %equations to determine the energy density of each one of the scalar %fields in terms of the density parameters and the density of the %standard components. It can be shown that the general solution is

%\begin{equation}
%    \rho_j = \frac{\Omega_j}{1- \sum^M_{j=1} \Omega_j} \left( %\rho_r + \rho_m + \rho_\Lambda \right) \, , \quad j=1, \ldots, M \, %.
%\end{equation}

\section{Stability Analysis of the System}

Here, we discuss the stability of the general phase space of the quintom models. To perform the stability analysis, we need to transfer the systems to a slightly different form by considering $\Omega_{\phi} ^{1/2} = r_{\phi}$ and $\Omega_{\sigma} ^{1/2} = r_{\sigma}$. This is done to avoid the singularity in the Jacobian matrix associated with the fixed points that have $\Omega_{\phi} = \Omega_{\sigma} = 0$. Hence, Eqs.~(\ref{eq:pol}) reduce to the following set of equations,

%\begin{widetext}
\begin{subequations}
\label{eq:pol_new}
  \begin{eqnarray}
    \theta_\phi^{\prime} &=& - 3 \sin \theta_\phi + y_{1\phi} \, , \label{eq:new_a} \\
    y_{1 \phi}^{\prime} &=&   3 /2 \gamma_{tot} y_{1 \phi}+ r_{\phi} \sin(\theta_\phi / 2) y_{2 \phi}  \, , \label{eq:new_b} \\ 
    r_\phi ^{\prime} &=& 3/2 (\gamma_{tot} - \gamma_{\phi}) r_\phi \, , \label{eq:new_c} \\
    \theta_\sigma ^{\prime} &=& - 3 \sinh \theta_\sigma - y_{1 \sigma}  \, , \label{eq:new_d} \\
    y_{1\sigma}^{\prime} &=&  3 /2 \gamma_{tot} y_{1 \sigma}+ r_{\sigma} \sinh(\theta_\sigma / 2) y_{2 \sigma}   \, , \label{eq:new_e} \\ 
    r_\sigma ^{\prime} &=& 3/2 (\gamma_{tot} - \gamma_{\sigma}) r_\sigma \, . \label{eq:new_f}
  \end{eqnarray}
\end{subequations}
%\end{widetext}

\subsection{Fixed Points and Eigenvalues}\label{appen:fixed_eigen}
The fixed points of the systems are listed in Table~\ref{tab:4}, and were obtained assuming that the contribution of the radiation component is negligible, $\Omega_r \ll 1$, so the total barotropic EOS is given by $\gamma_{tot} = 1 + (\gamma_{\phi} -1) r^2 _{\phi} + (\gamma_{\sigma} -1) r^2 _{\sigma}$. The values of the coefficients $A$ and $B$ are as follows:
\begin{subequations}
\begin{align}  
    A_{\pm} =& \frac{- \alpha_1 \pm \sqrt{\alpha_1 ^2 - 2 \alpha_0 (1 + 2 \alpha_2)}}{6 (1 + 2 \alpha_2)} \\
    B_{\pm} =& \frac{ \beta_1 \pm \sqrt{\beta_1 ^2 - 2 \beta_0 (1 + 2 \beta_2)}}{6 (1 + 2 \beta_2)}
\end{align}
\end{subequations}
The corresponding values of the cosmological parameters at the aforementioned fixed points are given in Table~\ref{tab:5}.

\begin{table}[]
\caption{Complete list of fixed points in the autonomous system~\eqref{eq:pol_new}. Fixed points with the name $m_i$ correspond to the matter dominated fixed point, $q_i$ corresponds to the quintessence dominated fixed point and $p_i$ for phatom domination.}
\label{tab:4}
\begin{tabular}{|c|c|c|c|c|}
\hline
\multicolumn{1}{|l|}{\begin{tabular}[c]{@{}l@{}}Fixed \\ Points\end{tabular}} & $r_{\phi_c}$ & $\sin \theta_{\phi_c}$ & $r_{\sigma_c}$ & $\sinh \theta_{\sigma_c}$ \\ \hline
$m_1$ & 0 & $n \pi$ & 0 & 0 \\ \hline
$m_2$ & 0 & $n \pi$ & 0 & $\sinh^2(\theta_{\sigma_c}  / 2) = \frac{\gamma_{tot}}{4 \beta_2}$ \\ \hline
$m_3$ & 0 & $\sin^2(\theta_{\phi_c}  / 2) =- \frac{\gamma_{tot}}{4 \alpha_2}$ & 0 & 0 \\ \hline
$m_4$ & 0 & $\sin^2(\theta_{\phi_c}  / 2) =- \frac{\gamma_{tot}}{4 \alpha_2}$ & 0 & $\sinh^2(\theta_{\sigma_c}  / 2) = \frac{\gamma_{tot}}{4 \beta_2}$ \\ \hline
$q_1$ & 1 & $n \pi$ & 0 & 0 \\ \hline
$q_2$ & 1 & $n \pi$ & 0 &   \multicolumn{1}{|l|}{\begin{tabular}[c]{@{}l@{}}$\sinh^2 (\theta_{\sigma c}/2) $ \\ $=\frac{1}{2\beta_2} \sin^2(\theta_{\phi c}/2)$\end{tabular}} \\ \hline
$q_3$ & 1 & \begin{tabular}[c]{@{}l@{}}$\sin(\theta_{\phi_c}  / 2) = A_{+}$\end{tabular} & 0 & 0 \\ \hline
$q_4$ & 1 & \begin{tabular}[c]{@{}l@{}}$\sin(\theta_{\phi_c}  / 2) = A_{-}$\end{tabular} & 0 & 0 \\ \hline
$q_5$ & 1 & \begin{tabular}[c]{@{}l@{}}$\sin(\theta_{\phi_c}  / 2) =A_{+}$\end{tabular} & 0 & \begin{tabular}[c]{@{}l@{}}$\sinh^2(\theta_{\sigma_c}  / 2) = \frac{{A^2_+}}{2 \beta_2}$\end{tabular} \\ \hline
$q_6$ & 1 & \begin{tabular}[c]{@{}l@{}}$\sin(\theta_{\phi_c}  / 2) =A_{-}$\end{tabular} & 0 & \begin{tabular}[c]{@{}l@{}}$\sinh^2(\theta_{\sigma_c}  / 2) = \frac{{A^2_-}}{2 \beta_2}$\end{tabular} \\ \hline
$p_1$ & 0 & $n \pi$ & 1 & 0 \\ \hline
$p_2$ & 0 & $n \pi$ & 1 & $\sinh(\theta_{\sigma c} /2 ) = B_+$ \\ \hline
$p_3$ & 0 & $n \pi$ & 1 & $\sinh(\theta_{\sigma c} /2 ) = B_-$ \\ \hline
$p_4$ & 0 & \begin{tabular}[c]{@{}l@{}}$\sin^2(\theta_{\phi_c}  / 2) =\frac{B^2 _+}{2 \alpha_2}$\end{tabular} & 1 & $\sinh(\theta_{\sigma c} /2 ) = B_+$ \\ \hline
$p_5$ & 0 & \begin{tabular}[c]{@{}l@{}}$\sin^2(\theta_{\phi_c}  / 2) =\frac{B^2 _-}{2 \alpha_2}$\end{tabular} & 1 & $\sinh(\theta_{\sigma c} /2 ) = B_-$ \\ \hline
\end{tabular}
\end{table}

\begin{table}[]
\caption{Value of different cosmological parameters; $\Omega_\phi, \Omega_\sigma, \gamma_\phi, \gamma_\sigma, \gamma_{tot}$  corresponding to different fixed points given in Table~\ref{tab:4}.}
\label{tab:5}
\begin{tabular}{|c|c|c|c|c|c|}
\hline
\multicolumn{1}{|l|}{\begin{tabular}[c]{@{}l@{}}Fixed \\ \\ Points\end{tabular}} & $\Omega_\phi$ & $\gamma_\phi$ & $\Omega_\sigma$ & $\gamma_\sigma$ & $\gamma_{tot}$ \\ \hline
$m_1$ & \multicolumn{1}{c|}{0} & \begin{tabular}[c]{@{}l@{}}0; for $n$ even\\ 2; for $n$ odd\end{tabular} & 0 & 0 & 1 \\ \hline
$m_2$ & \multicolumn{1}{c|}{0} & \begin{tabular}[c]{@{}l@{}}0; for $n$ even\\ 2; for $n$ odd\end{tabular} & 0 & $-\frac{\gamma_{tot}}{2 \beta_2}$ & 1 \\ \hline
$m_3$ & 0 & $-\frac{\gamma_{tot}}{2 \alpha_2}$ & 0 & 0 & 1 \\ \hline
$m_4$ & 0 & $-\frac{\gamma_{tot}}{2 \alpha_2}$ & 0 & $-\frac{\gamma_{tot}}{2 \beta_2}$ & 1 \\ \hline
$q_1$ & \multicolumn{1}{c|}{1} & \begin{tabular}[c]{@{}l@{}}0; for $n$ even\\ 2; for $n$ odd\end{tabular} & 0 & 0 & $\gamma_\phi$ \\ \hline
$q_2$ & \multicolumn{1}{c|}{1} & \begin{tabular}[c]{@{}l@{}}0; for $n$ even\\ 2; for $n$ odd\end{tabular} & 0 & $- \frac{\gamma_\phi}{2 \beta_2}$ & $\gamma_\phi$ \\ \hline
$q_3,q_4$ & \multicolumn{1}{c|}{1} & $2 A^2 _\pm$ & 0 & 0& $\gamma_\phi$ \\ \hline
$q_5,q_6$ & \multicolumn{1}{c|}{1} & $2 A^2 _\pm$ & 0 & $- \frac{A^2 _\pm}{\beta_2}$& $\gamma_\phi$ \\ \hline
$p_1$ & \multicolumn{1}{c|}{0} & \begin{tabular}[c]{@{}l@{}}0; for $n$ even\\ 2; for $n$ odd\end{tabular} & 1 & 0 & $\gamma_\sigma$ \\ \hline
$p_2,p_3$ & \multicolumn{1}{c|}{0} & \begin{tabular}[c]{@{}l@{}}0; for $n$ even\\ 2; for $n$ odd\end{tabular} & 1 & $-2 B^2 _\pm$ & $\gamma_\sigma$ \\ \hline
$p_4, p_5$ & \multicolumn{1}{c|}{0} & $\frac{B^2 _\pm}{2 \alpha_2}$ & 1 & $-2 B^2 _\pm$ & $\gamma_\sigma$ \\ \hline
\end{tabular}
\end{table}

\begin{table*}[]
\caption{The table displays eigenvalues for fixed points listed in Table~\ref{tab:4}.}
\label{tab:6}
\begin{tabular}{|c|l|}
\hline
\multicolumn{1}{|l|}{\begin{tabular}[c]{@{}l@{}}Fixed Points\end{tabular}} & \multicolumn{1}{c|}{Eigenvalues} \\ \hline
$m_1$ & $ -3, -3, \frac{3}{2}, \frac{3}{2}, \frac{3}{2}, \frac{3}{2} $\ \\ \hline
$m_2$ & \begin{tabular}[c]{@{}l@{}}$-3,\frac{3}{2},\frac{3}{2},\frac{3}{2} \left(1+\frac{1}{2 \beta _2}\right),\frac{3 \left(-1-3 \beta _2-\sqrt{1+2 \beta _2-7 \beta _2^2}\right)}{4 \beta _2},\frac{3 \left(-1-3 \beta _2+\sqrt{1+2 \beta _2-7 \beta _2^2}\right)}{4 \beta _2}$\end{tabular} \\ \hline
$m_3$ & $ -3,\frac{3}{2},\frac{3}{2},\frac{3}{2} \left(1-\frac{1}{2 \alpha _2}\right),\frac{3 \left(1+\alpha _2-\sqrt{1-10 \alpha _2+33 \alpha _2^2}\right)}{4 \alpha _2},\frac{3 \left(1+\alpha _2+\sqrt{1-10 \alpha _2+33 \alpha _2^2}\right)}{4 \alpha _2}$ 
%$\frac{3}{4} \left(-1-\beta _2-2 \beta _2^2+\sqrt{9+6 \beta _2-3 \beta _2^2-4 \beta _2^3+4 \beta _2^4}\right)$\end{tabular} 
\\ \hline
$m_4$ & $0,0,-\frac{3}{2}(2 + \frac{1}{\alpha _2}),\frac{3}{4} \left(2+\frac{1}{\alpha _2}\right),-\frac{3 \left(1+3 \beta _2+\sqrt{1+2 \beta _2+9 \beta _2^2}\right)}{4 \beta _2},\frac{3 \left(-1-3 \beta _2+\sqrt{1+2 \beta _2+9 \beta _2^2}\right)}{4 \beta _2}$ \\ \hline
$q_1, q_2$ & $-3,-3,0,0,\frac{1}{2} \left(-3-\sqrt{9+2 \alpha _0}\right),\frac{1}{2} \left(-3+\sqrt{9+2 \alpha _0}\right)$
%\\ $\frac{3}{4} \left(-1-\beta _2-2 \beta _2^2+\sqrt{9+6 \beta _2-3 \beta _2^2-4 \beta _2^3+4 \beta _2^4}\right)$\end{tabular} 
\\ \hline
$q_3, q_4$ & $-3,3 A_{\pm}^2,3A_{\pm}^2,-3 \left(1-2 A_{\pm}^2\right), Q_1, Q_2$ \\ \hline
$q_5, q_6$ & $-3 \left(1-2 A_{\pm}^2\right), \frac{3}{2} \left(2 A_{\pm}^2+\frac{A_{\pm}}{\beta _2}\right), Q_1, Q_2,Q_3, Q_4$ \\ \hline
$p_1$ & $-3,-3,0,0,\frac{1}{2} \left(-3-\sqrt{9-2 \beta _0}\right),\frac{1}{2} \left(-3+\sqrt{9-2 \beta _0}\right)$ \\ \hline
$p_2,p_3$ & $-3,-3 B_{\pm}^2,-3 B_{\pm}^2,-3 \left(1+2 B_{\pm}^2\right), P_1, P_2$ \\ \hline
$p_4,p_5$ & $-3 \left(1+2 B_{\pm}^2\right),\frac{3}{2} \left(-2 B_{\pm}^2-\frac{B_{\pm}}{\alpha _2}\right), P_1, P_2, P_3, P_4$ \\ \hline
\end{tabular}
\end{table*}

 Fixed points can be classified into three different classes: i) matter dominated, ii) quintessence dominated, and iii) phantom dominated (see Table~\ref{tab:5}).  Fixed points dominated by matter are represented by $m_i$, points dominated by quintessence are represented by $q_i$, and points dominated by phantoms are represented by $p_i$. 

 \subsection{Eigenvalues}
\label{appen:eigen}
 The eigenvalues of these fixed points are listed in Table~\ref{tab:6}. To find them, we have considered $n$ to be even for the fixed points where $\sin\theta_{\phi c} = n \pi$, and the corresponding $\gamma_\phi = 0$ that is of physical interest. Odd-$n$ leads to $\gamma_\phi = 2$ in which the quintessence field behaves as a stiff fluid. For the complete expressions of $Q_1, Q_2, Q_3, Q_4, P_1, P_2, P_3, P_4$ see Sec.~\ref{appen:eigen}. 
 
 Certain fixed points may not be hyperbolic, meaning that some of their eigenvalues are zero. This means that the standard linear stability analysis cannot be used to determine their stability. If one or more of the non-zero eigenvalues are positive, then the center manifold theorem implies that the fixed point is likely to be unstable. If all the non-zero eigenvalues are negative, further investigation is necessary to determine the stability of the fixed point.

\subsection{Stability of the matter dominated points}
The points $m_1$ to $m_3$ have a positive eigenvalue, making them unstable fixed points. The point $m_4$ is a nonhyperbolic fixed point, as it has zero eigenvalues. It can be easily seen that for the fixed points mentioned, all eigenvalues cannot be simultaneously negative for any combination of $\alpha_2$ and $\beta_2$. Consequently, they are also unstable.

\subsection{Stability of the quintessence dominated points}
The fixed points $q_1$ and $q_2$ may be late-time attractors, as all their eigenvalues, apart from zero, can be negative for $-9/2\leq \alpha_0 < 0$. On the other hand, $q_3$ and $q_4$ are unstable, as they have positive eigenvalues. It is difficult to make a general statement about the stability of $q_5$ and $q_6$, as the forms of some of their eigenvalues are very complex. The stability of these fixed points is dependent on the selection of the parameters $\alpha$ and $\beta$ in a complicated manner.

\subsection{Stability of the phantom dominated points}
The fixed point $p_1$ may have all negative eigenvalues in the range of $0<\beta _0\leq 9/2$, making it stable in this particular interval. The eigenvalues of the fixed points $p_2$ to $p_5$ are complex functions of the parameters $\alpha$ and $\beta$, making their stability difficult to determine.  

\begin{widetext}
\begin{multline}
Q_{1,2}  =  \frac{1}{4} \left(-6+18 A_{\pm}^2+2 A_{\pm} \alpha _1+24 A_{\pm}^2 \alpha _2 \right)  \\  \pm \frac{1}{4} \left( \sqrt{\left(6-18 A_{\pm}^2-2 A_{\pm} \alpha _1-24 A_{\pm}^2 \alpha _2\right){}^2-8 \left(-54 A_{\pm}^2+72 A_{\pm}^4-\alpha _0+2 A_{\pm}^2 \alpha _0-12 A_{\pm} \alpha _1+18 A_{\pm}^3 \alpha _1-108 A_{\pm}^2 \alpha _2+144 A_{\pm}^4 \alpha _2\right)}\right)  
\end{multline}

\begin{multline}
Q_{3,4}  = -\frac{3 A_{\pm}+3 \beta _2+6 A_{\pm} \beta _2-3 A_{\pm}^2 \beta _2 \pm \sqrt{\left(-3 A_{\pm}-3 \beta _2-6 A_{\pm} \beta _2+3 A_{\pm}^2 \beta _2\right){}^2+4 \beta _2 \left(-18 A_{\pm}^2+9 A_{\pm}^3-27 A_{\pm} \beta _2+9 A_{\pm}^2 \beta _2\right)}}{2 \beta _2}
\end{multline}

\begin{multline}
P_{1,2}  =  \frac{1}{4} \left(-6-18 B_{\pm}^2+2 B_{\pm} \beta _1-24 B_{\pm}^2 \beta _2 \right) \\ \pm \frac{1}{4} \left( \sqrt{\left(6+18 B_{\pm}^2-2 B_{\pm} \beta _1+24 B_{\pm}^2 \beta _2\right){}^2-8 \left(54 B_{\pm}^2+72 B_{\pm}^4+\beta _0+2 B_{\pm}^2 \beta _0-12 B_{\pm} \beta _1-18 B_{\pm}^3 \beta _1+108 B_{\pm}^2 \beta _2+144 B_{\pm}^4 \beta _2\right)}\right)
\end{multline}

\begin{multline}
P_{3,4}  = \frac{3 B_{\pm}-3 \alpha _2+6B_{\pm} \alpha _2-3 B_{\pm}^2 \alpha _2 \pm \sqrt{4 \alpha _2 \left(-18 B_{\pm}^2+9 B_{\pm}^3+27 B_{\pm} \alpha _2-9 B_{\pm}^2 \alpha _2\right)+\left(3 B_{\pm}-3 \alpha _2+6 B_{\pm} \alpha _2-3 B_{\pm}^2 \alpha _2\right){}^2}}{2 \alpha _2}
\end{multline}
\end{widetext}

\subsection{General discussion on stability}
The stability of fixed points is contingent on the selection of active parameters $\alpha, \beta$. A qualitative analysis of the entire phase space of the system was conducted, which revealed that all matter-dominated fixed points are unstable, suggesting that the universe may have gone through some of these fixed points. The probable late-time attractors of the universe are either quintessence-dominated or phantom-dominated. However, the analysis of a general case is complicated due to the dependency of the eigenvalues on the values of $\alpha$ and $\beta$. To simplify this, a particular functional form of the quintessence and phantom potentials could be chosen, which would fix the values of their active parameters and eigenvalues. This would enable a detailed study of the dynamics of any particular case of interest.

\bibliography{quintom}

\end{document}